\documentclass[prl,aps,twocolumn,floats,showpacspsfig]{revtex4}
\usepackage{amsmath,amssymb}
\usepackage{amsfonts}
\usepackage{mathrsfs}
\usepackage{epic}
\usepackage{eepic}
\usepackage{enumerate}
\usepackage{graphicx}
\usepackage[hang,small,bf%,nooneline
]{caption}
\newcounter{app}
\newcounter{sapp}[app]

\newcommand{\ds}{\displaystyle}
\def\EXP{\textrm{{\large e}}}
\newcommand{\qs}{\mathsf{q}}
\newcommand{\bs}{\mathsf{b}}
\newcommand{\ii}{\mathsf{i}}

\renewcommand{\t}{\theta}
\newcommand{\crs}{\eta}
\newcommand{\norma}{F}
\newcommand{\Gfun}{\Phi}
\newcommand{\be}{=}

\renewcommand{\Im}{\mathrm{Im}}
\renewcommand{\Re}{\mathrm{Re}}
\def\bea{\begin{eqnarray}}
\def\eea{\end{eqnarray}}

\begin{document}

\title{Exact solution of the Faddeev-Volkov model}

\author{Vladimir V. Bazhanov}
\author{Vladimir V. Mangazeev}
\author{Sergey M. Sergeev}

\affiliation{Department of Theoretical Physics,\\
         Research School of Physical Sciences and Engineering,\\
    Australian National University, Canberra, ACT 0200, Australia.}

\begin{abstract}
The Faddeev-Volkov model is an Ising-type lattice model with positive
Boltzmann weights  where the spin variables take continuous values
on the real line. It serves as a lattice analog of the
$\sinh$-Gordon and Liouville models and intimately
connected with the modular double of the quantum group
$U_q(sl_2)$. The free energy of the model is exactly
calculated in the thermodynamic limit.
In the quasi-classical limit $c\to+\infty$ the model describes
quantum fluctuations of
discrete conformal transformations connected
with the Thurston's discrete analogue of the Riemann mappings
theorem. In the strongly-coupled limit $c\to 1$ the model turns
into a discrete version of the ${\mathcal D}=2$ Zamolodchikov's
``fishing-net'' model.
\end{abstract}

\maketitle
Faddeev and Volkov \cite{Volkov:1992,FV:1993,Faddeev:1994} obtained a very
interesting solution of the star-triangle relation.
There is only a few two-dimensional solvable lattice models \cite{Baxterbook}
where the Yang-Baxter equation takes its distinguished
``star-triangular'' form.
These models include the Ashkin-Teller \cite{Ashkin:1943},
Kashiwara-Miwa \cite{Kashiwara:1986} and chiral Potts \cite{Baxter:1987eq}
models (the
latter also contains the Ising, self-dual Potts \cite{Potts:1952}
and Fateev-Zamolodchikov \cite{FZ82}  models as particular cases).
All above models are also distinguished by a positivity of
Boltzmann weights --- the property that is naturally expected for physical
applications, but rarely realized for generic solutions of the
Yang-Baxter equation. Recently, we have observed \cite{BMS07a} that the
Faddeev-Volkov model \cite{Volkov:1992,FV:1993,Faddeev:1994}
also possesses the positivity property. Apart from being an interesting
model of statistical mechanics
and quantum field theory in its own rights (it serves as a lattice
version of the sinh-Gordon and Liouville models
\cite{FKV:2001,Bytsko:2006}) this model
has remarkable connections with discrete geometry. As shown in
\cite{BMS07a} it describes quantum fluctuations of circle patterns
\cite{BSp} and associated discrete conformal transformations connected
with with Thurston's discrete analogue of the Riemann mapping
theorem \cite{St1}.
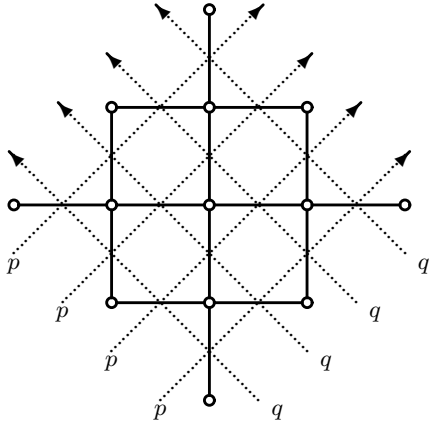
\begin{figure}[th]
\begin{center}
\setlength{\unitlength}{0.065mm}
\begin{picture}(900,850)(50,75)
\Thicklines
% rapidities
\dottedline[$\cdot$]{12}(600,100)(100,600)\put(100,600){\vector(-1,1){10}}
\put(610,70){ $q$}
\dottedline[$\cdot$]{12}(700,200)(200,700)\put(200,700){\vector(-1,1){10}}
\put(710,170){ $q$}
\dottedline[$\cdot$]{12}(800,300)(300,800)\put(300,800){\vector(-1,1){10}}
\put(810,270){ $q$}
\dottedline[$\cdot$]{12}(900,400)(400,900)\put(400,900){\vector(-1,1){10}}
\put(910,370){ $q$}
\dottedline[$\cdot$]{12}(100,400)(600,900)\put(600,900){\vector(1,1){10}}
\put(70,370){ $p$}
\dottedline[$\cdot$]{12}(200,300)(700,800)\put(700,800){\vector(1,1){10}}
\put(170,270){ $p$}
\dottedline[$\cdot$]{12}(300,200)(800,700)\put(800,700){\vector(1,1){10}}
\put(270,170){ $p$}
\dottedline[$\cdot$]{12}(400,100)(900,600)\put(900,600){\vector(1,1){10}}
\put(370,70){ $p$}
% spins
\put(500,900){\circle{20}}
\put(300,700){\circle{20}}\put(500,700){\circle{20}}\put(700,700){\circle{20}}
\put(100,500){\circle{20}}\put(300,500){\circle{20}}\put(500,500){\circle{20}}
\put(700,500){\circle{20}}\put(900,500){\circle{20}}
\put(300,300){\circle{20}}\put(500,300){\circle{20}}\put(700,300){\circle{20}}
\put(500,100){\circle{20}}
% edges
%\allinethickness{.4mm}
%
\path(500,110)(500,290)\path(500,310)(500,490)\path(500,510)(500,690)\path(500,710)(500,890)
\path(300,310)(300,490)\path(300,510)(300,690)
\path(700,310)(700,490)\path(700,510)(700,690)

\path(110,500)(290,500)\path(310,500)(490,500)\path(510,500)(690,500)\path(710,500)(890,500)
\path(310,300)(490,300)\path(510,300)(690,300)
\path(310,700)(490,700)\path(510,700)(690,700)
% shade
%\allinethickness{0.1\unitlength}
%\put(400,300){\shde}\put(400,500){\shde}\put(600,300){\shde}\put(600,500){\shde}
% spins
\end{picture}
%\end{picture}
\end{center}
\caption{The square lattice (solid lines) and its medial ``rapidity''
lattice (dashed lines).}\label{fig1}
\end{figure}

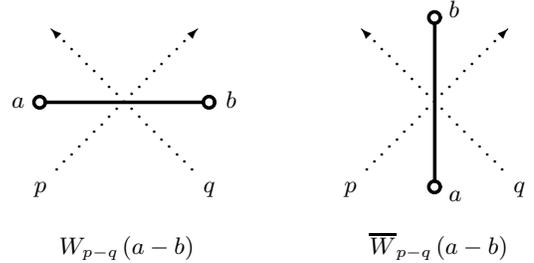
\begin{figure}[thb]
\begin{center}
\setlength{\unitlength}{0.75mm}
\begin{picture}(100,43)(-10,5)
\put(0,15){\begin{picture}(30,30) \thinlines
\dottedline[$\cdot$]{2}(3,3)(27,27)\put(27,27){\vector(1,1){1}}
\dottedline[$\cdot$]{2}(27,3)(3,27)\put(3,27){\vector(-1,1){1}}
\put(-1,-1){$p$}\put(29,-1){$q$} \allinethickness{.5mm}
\put(0,15){\circle{2}}\put(30,15){\circle{2}}\path(1,15)(29,15)
\put(-5,14){$a$}\put(33,14){$b$} \put(2,-12){ $W_{p-q\,}(a-b)$}
\end{picture}}
\put(55,15){\begin{picture}(30,30) \thinlines
\dottedline[$\cdot$]{2}(3,3)(27,27)\put(27,27){\vector(1,1){1}}
\dottedline[$\cdot$]{2}(27,3)(3,27)\put(3,27){\vector(-1,1){1}}
\put(-1,-1){$p$}\put(29,-1){$q$} \allinethickness{.5mm}
\put(15,0){\circle{2}}\put(15,30){\circle{2}}\path(15,1)(15,29)
\put(17.5,30){$b$}\put(17.5,-2.5){$a$} \put(2,-12){
$\overline{W}_{p-q\,}(a-b)$}
\end{picture}}
\end{picture}
\end{center}
\caption{Two types of Boltzmann weights.}\label{fig2}
\end{figure}

Consider the square lattice, shown in Fig.\ref{fig1}.
Each site $i$ of a lattice is assigned with a spin variable
$\sigma_i\in {\mathbb R}$, taking \emph{continuous real values}.
Two spins $a$ and $b$ interact only if they are adjacent
(connected with an edge of the lattice).
Typical
horizontal and vertical edges are shown in Fig.\ref{fig2}.
The corresponding Boltzmann weights are parametrized through the
so-called ``rapidity variables'' $p$ and $q$ associated with the
oriented dashed lines in Fig.\ref{fig1}. In our case the weights
depends only on the spin and rapidity differences $a-b$ and $p-q$,
where $a$ and $b$ are the spins at the ends of the edge.
We will denote them as $W_{p-q}(a-b)$ and $\overline{W}_{p-q}(a-b)$
for the horizontal and vertical edges, respectively.
The partition function is defined as
\begin{equation}
Z=\int  \prod_{(i,j)}
W_{p-q}(\sigma_i-\sigma_j)\ \prod_{(k,l)}
\overline{W}_{p-q}(\sigma_k-\sigma_l)\ \prod_i d\sigma_i\;.\label{Z-def}
\end{equation}
where the first product is over all horizontal edges $(i, j)$ and the
second is over all vertical edges $(k,l)$. The integral is taken over
all the interior spins; the boundary spins are kept fixed.
The weights $W_{p-q}(a-b)$ and $\overline{W}_{p-q}(a-b)$
are related with non-compact representations \cite{Ponsot-1999} of the
modular double \cite{Faddeev:1999} of the quantum group
$U_\qs({sl}_2)\otimes U_{\tilde{\qs}}({sl}_2)$, where
%\begin{equation}
$\qs=e^{i\pi { \bs}^2}$ and $\tilde{\qs}=e^{-i\pi/ {\bs}^2}$.
%\end{equation}
The modular parameter
$\bs$ is connected to the Liouville central charge
$c_L=1+6\,(\bs+\bs^{-1})^2$. It is convenient to define
\begin{equation}
\eta=\frac{1}{2}(\bs+\bs^{-1}),
\qquad \overline{\qs}
=\ii\,\exp\left(\frac{\ii\pi(\bs-\bs^{-1})}{2(\bs+\bs^{-1})}\right)\;.\label{qbar}
\end{equation}
The main physical regimes of the model
\begin{equation}
{\rm (i)}\ \bs>0,\qquad {\rm and \ \ \ \ (ii)}\  |\bs|=1, \qquad
\Im(\bs^2)>0.\label{regimes}
\end{equation}
correspond to real values of $\eta$.
For the regime (i) it is enough to consider the range $0<\bs\le1$
(due to the symmetry $\bs\leftrightarrow \bs^{-1}$).

Introduce the non-compact
quantum dilogarithm \cite{Faddeev:1994},
\bea
\varphi(z)&\be&\exp\left(\ds \frac{1}{4}\int_{\mathbb{R}+\ii 0}
\frac{\EXP^{-2\ii zw}}{\textrm{sinh}(w\bs)\textrm{sinh}(w/\bs)}\
\frac{dw}{w}\right)\nonumber\\
&&\nonumber\\
&=&\ds
\frac{(-\qs\,e^{2\pi z\,\bs}\,;\ \qs^2)_\infty} {(
-\tilde \qs\,e^{2\pi z\, \bs^{-1}};\tilde
\qs^{\,2})_\infty}\;,\label{fi-def}
\eea
where $\ds (x,\qs)_\infty=\prod\nolimits_{k=0}^\infty (1-\qs^k x)$.
The product representation above is valid for $\Im (\bs^2)>0$.

The Boltzmann weights $W_{\theta\,}(s)$ and
$\overline{W}_{\t\,}(s)$ are defined as
\begin{equation}\label{W}
W_{\t\,}(s)\be\frac{1}{\norma_\t}\ \EXP^{2\crs \t s}\
\frac{\varphi(s+\ii \crs \t/\pi)}{\varphi(s-\ii \crs \t/\pi)}\;,\quad
\overline{W}_{\t\,}(s)\be W_{\pi-\t}(s)\;,
\end{equation}
where $\t$ and $s$ stand for the rapidity and spin differences,
respectively. The normalization factor $\norma_\t$ has the form
\begin{equation}\label{F}
\norma_\t\; \be \;\EXP^{\ii\crs^2 \t^2/\pi + \ii \pi (1-8\crs^2)/24}\
\Gfun(2\ii \crs\/ \t/\pi)\;.
\end{equation}
where
\bea
\Gfun(z)\ds &\be& \exp\Big(\frac{1}{8}\int\nolimits\limits_{\mathbb{R}+\ii
0} \frac{\EXP^{-2\ii z w}} {\sinh(w\bs)\sinh(w/\bs)\cosh(2w\crs)}\
\frac{dw}{w}\Big)\nonumber\\
&&\nonumber\\
&=&\ds\frac{(\qs^2 e^{2\pi
z\,\bs};\qs^4)_\infty} {(\tilde{\qs}^{\,2}e^{2\pi z\,
\bs^{-1}};\tilde{\qs}^{\,4})_\infty} \frac{\ds(-\overline{\qs} e^{{\pi
z}/(2\eta)}; \overline{\qs}^{\,2})_\infty} {\ds(\overline{\qs}
\,e^{{\pi z}/(2\eta)};\overline{\qs}^{\,2})_\infty}
\eea
with $\overline{\qs}$ defined in \eqref{qbar}.
The main properties of $\varphi(z)$ and $\Gfun(z)$
are discussed in \cite{BMS07a}.

When the parameter $\bs$ belongs to either of the regimes
\eqref{regimes}, the Boltzmann weights
$W_{\t\,}(s)$ and $\overline{W}_\t(s)$  are real and positive
for $0<\theta<\pi$ and real $s$.
They are even functions of
the variable $s$ and decay exponentially,
$W_{\t\,}(s)\; \simeq \; \norma_\t^{-1}\EXP^{-2\crs \t |s|}$, when
$s\to \pm\infty$. The weight $W_{\t\,}(s)$, considered as a function
of $\theta$ at fixed real $s$, is analytic and non-zero
in the strip $0<\Re\,\theta<\pi$  .

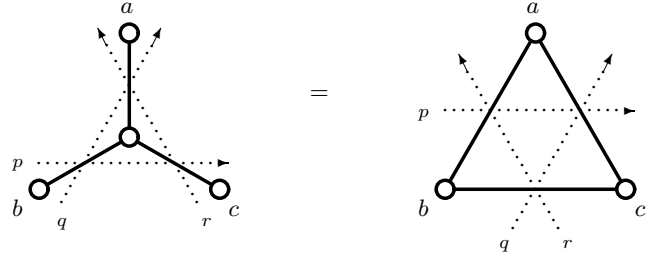
\begin{figure}[thb]
\begin{center}
\setlength{\unitlength}{1.2mm}
\begin{picture}(65,26)
\put(0,5){\begin{picture}(20,20) \thinlines
\dottedline[$\cdot$]{1}(2.5,-1.44)(12.5,15.877)\put(12.5,15.877){\vector(1,2){1}}
\put(2,-4){\scriptsize $q$}
\dottedline[$\cdot$]{1}(17.5,-1.44)(7.5,15.877)\put(7.5,15.877){\vector(-1,2){1}}
\put(18,-4){\scriptsize $r$}
\dottedline[$\cdot$]{1}(0,2.807)(20,2.887)\put(20,2.887){\vector(1,0){1}}
\put(-3,2.5){\scriptsize $p$} \allinethickness{.5mm}
\put(0,0){\circle{2}}\put(20,0){\circle{2}}\put(10,17.32){\circle{2}}
\put(-3,-3){$b$}\put(21,-3){$c$}\put(9,19.5){$a$}
\put(10,5.7735){\circle{2}}\path(0.866,0.5)(9.134,5.27)
\path(10,6.77)(10,16.32)\path(10.866,5.2735)(19.134,0.5)
\end{picture}}
\put(30,15){$=$}
\put(45,5){\begin{picture}(20,20) \thinlines
\dottedline[$\cdot$]{1}(12.5,-4.33)(2.5,12.99)\put(2.5,12.99){\vector(-1,2){1}}
\put(13,-6.5){\scriptsize $r$}
\dottedline[$\cdot$]{1}(7.5,-4.33)(17.5,12.99)\put(17.5,12.99){\vector(1,2){1}}
\put(6,-6.5){\scriptsize $q$}
\dottedline[$\cdot$]{1}(0,8.66)(20,8.66)\put(20,8.66){\vector(1,0){1}}
\put(-3,8){\scriptsize $p$} \allinethickness{.5mm}
\put(0,0){\circle{2}}\put(20,0){\circle{2}}\put(10,17.32){\circle{2}}
\put(-3,-3){$b$}\put(21,-3){$c$}\put(9,19.5){$a$}
\path(0.5,0.866)(9.5,16.45)\path(1,0)(19,0)
\path(10.5,16.45)(19.5,0.866)
\end{picture}}
\end{picture}
\end{center}
\caption{Star-triangle relation.}\label{fig3}
\end{figure}

The weights $W_{\t}$ and $\overline{W}_{\t}$ satisfy several
functional relations: the duality relation
\begin{equation}\label{Wp}
\overline{W}_{\t\,}(s)\;=\;\int_\mathbb{R}\  dx\ \EXP^{2\pi\ii xs}
\ W_{\t\,}(x)\;,
\end{equation}
the inversion relations,
\bea
\lim_{\varepsilon\to
0^+}\int_{\mathbb{R}} \; dc\; \overline{W}_{\ii
t+\varepsilon\,}(a-c) \overline{W}_{-\ii t
+\varepsilon\,}(c-b)&=&\delta(a-b)\;,\nonumber\\
&&\nonumber\\
W_{\t\,}(a-b) W_{-\t\,}(a-b)&=&1\;,\label{invrel}
\eea
where $t$ is real,  and
the star-triangle relation, Fig.\ref{fig3}:
\bea
\int_\mathbb{R} d\sigma \ \overline{W}_{ q-r\,}(a-\sigma) \ W_{
p-r\,}(c-\sigma) \ \overline{W}_{ p-q\, } (\sigma-b)&&\nonumber\\
&&\nonumber\\
=W_{p-q\,}(c-a) \overline{W}_{ p-r\,}(a-b) \ W_{q-r\,}(c-b)\;.&&\label{str}
\eea
%These relations can proven by means of the
%residue theorem for $\Im(\bs^2)>0$ and then analytically continued
%to the case of real $\bs$.

We used the inversion relation method \cite{Str79,Zam79,Bax82inv}
to exactly calculate the partition function \eqref{Z-def} in the
thermodynamic limit. The result is included in the normalization
of the of Boltzmann weights, so that the free energy per edge,
\begin{equation}
\beta f_{edge}=-\lim_{N\to \infty}{N^{-1}\log Z}=0\ , \label{fzero}
\end{equation}
vanishes when the number of edges, $N$, tends to infinity (provided the
boundary spins are kept finite).
The weights \eqref{W} attain their maximal values at $s=0$.
In the quasi-classical regime (i),
$0<\bs<1$, the value of the constant
$W_{\frac{\pi}{2}}(0)$ slowly interpolates between the values
\begin{equation}
W_{\frac{\pi}{2}}(0)\Big|_{\bs=0}=\EXP^{\frac{G}{\pi}},\qquad
W_{\frac{\pi}{2}}(s)\Big|_{\bs=1}=\sqrt{2},
\end{equation}
where $G=0.915965\ldots$ is the Catalan's constant.
The $s$-dependence of the
two-spin interaction energy $E(s)=-\log W_{\frac{\pi}{2}}(s)$
is quadratic, $E(s)-E(0)\simeq s^2$,
for small $s$ and gradually becomes
linear, $E(s)\simeq \pi \eta|s|$, for large $s$.

\begin{figure}[ht]
\setlength{\unitlength}{0.16mm}
\begin{picture}(250,250)
\put(25,25){\begin{picture}(200,200) \thinlines
% rapidities
% \put(0,50){\vector(1,1){150}}\put(50,0){\vector(1,1){150}}
% \put(150,0){\vector(-1,1){150}}\put(200,50){\vector(-1,1){150}}
 \Thicklines
 \put(0,100){\circle{10}}\put(95,-20){\scriptsize $\rho_S$}
 \put(100,100){\circle{10}}\put(110,110){\scriptsize $\rho$}
 \put(200,100){\circle{10}}\put(210,95){\scriptsize $\rho_E$}
 \put(100,0){\circle{10}}\put(-40,95){\scriptsize $\rho_W$}
 \put(100,200){\circle{10}} \put(95,220){\scriptsize $\rho_N$}
% palki
 \path(5,100)(95,100)\path(105,100)(195,100)
\path(100,5)(100,95)\path(100,105)(100,195)
 \put(50,150){\circle*{10}}\put(40,165){\scriptsize $\psi_{NW}$}
 \put(150,150){\circle*{10}}\put(140,165){\scriptsize $\psi_{NE}$}
 \put(50,50){\circle*{10}}\put(40,30){\scriptsize $\psi_{SW}$}
 \put(150,50){\circle*{10}}\put(140,30){\scriptsize $\psi_{SE}$}
\end{picture}}
\end{picture}\caption{Arrangement of the $\rho$- and $\psi$-variables
 around a four-edge star.}\label{fig4}
\end{figure}
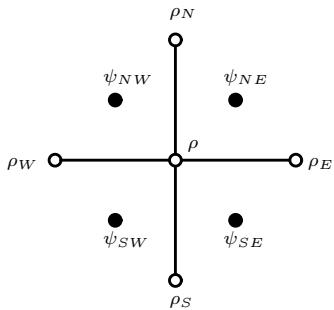

The parameter $\bs^2$ plays the role of the Planck constant.
The quasi-classical limit $\bs\to0$ (where
the model reveals remarkable connections with the discrete geometry)
was considered in \cite{BMS07a}. It is worth
mentioning some details here. When $\bs\to0$ the weight
function $W_\theta(s)$ acquires a very narrow bell-shaped form and rapidly
decays outside the small interval $|s|<\bs/\theta$. The partition
function \eqref{Z-def} can be then calculated with saddle point
method. It is convenient to define $\theta=p-q$ and
use rescaled spin variables $\rho=\{\rho_1,\rho_2,\ldots\}$, \
given by $\rho_i=2\pi \bs\sigma_i$.
The leading asymptotics of \eqref{W} reads
\begin{equation}
W_{\theta}({\rho}/{2\pi \bs})
=e^{ -A(\theta\/|\rho)/{2\pi \bs^2}  +
O(\bs^0)}\;,\quad \bs\to 0\ , \label{asyexp}
\end{equation}
where the function
\begin{equation}
A(\theta\/|\rho)=\frac{1}{\ii} \int_0^\rho
\log f_\theta(\xi)\, d\xi\ ,\quad
f_\theta(\xi)=\frac{1+\EXP^{\xi + \ii \theta}}{\EXP^{\xi} +
\EXP^{\ii \theta}}\label{A-def}
\end{equation}
is simply related to the Euler dilogarithm. The stationary point
of the integral \eqref{Z-def} is determined by the classical equation
of motion
\begin{equation}
f_\theta(\rho_W-\rho)
f_{\pi-\theta}(\rho_N-\rho)
f_\theta(\rho_E-\rho)
f_{\pi-\theta}(\rho_S-\rho)=1, \label{cr-eq}
\end{equation}
connecting five neighboring spins $\rho_W,\rho_N,\rho_E,\rho_S$ and
$\rho$, arranged as in Fig.\ref{fig4}
This non-linear
difference equation is a particular (square lattice) case of the
``cross-ratio equations'' \cite{NC,BoSur}. The latter can be viewed as
discrete analogues of the Cauchy-Riemann conditions for analytic
functions. In particular, Eq.\eqref{cr-eq} defines
discrete conformal transformations \footnote{These transformations are nicely visualized by the circle patterns
with prescribed intersection angles \cite{BSp}
 The (rescaled) spin variables,
solving \eqref{cr-eq}, define the radii $r_i=e^{\rho_i}$ of the
circles, while the rapidity variable $\theta$ defines the circle
intersection angles, see \cite{BMS07a} for more details.}
of the square lattice \cite{S,BoSur}, where the
local dilatation factors $r_i=\exp(\rho_i)$ are determined by
the rescaled spin variables $\rho_i$,
solving \eqref{cr-eq}.
\begin{figure}[ht]
\setlength{\unitlength}{0.15mm}
\begin{picture}(450,200)
\put(250,0){\begin{picture}(200,200) \thinlines
% rapiditie
% \put(20,20){\vector(1,1){160}}
% \put(180,20){\vector(-1,1){160}}
 \Thicklines
 \put(100,30){\circle{10}}
 \put(100,170){\circle{10}}
 \path(100,35)(100,165)
 \put(30,100){\circle*{10}}
 \put(170,100){\circle*{10}}
\thinlines
 \dashline[40]{10}(95,35)(35,95)
 \dashline[40]{10}(35,105)(95,165)
 \dashline[40]{10}(105,165)(165,105)
 \dashline[40]{10}(105,35)(165,95)
 \put(95,0){\scriptsize $\rho_S$}
 \put(95,190){\scriptsize $\rho_N$}
 \put(-5,95){\scriptsize $\psi_W$}
 \put(190,95){\scriptsize $\psi_E$}
% \put(110,95){\scriptsize $\theta$}
%
\end{picture}}
\put(0,0){\begin{picture}(200,200) \thinlines
% rapidities
% \put(20,20){\vector(1,1){160}}
% \put(180,20){\vector(-1,1){160}}
 \Thicklines
 \put(30,100){\circle{10}}
 \put(170,100){\circle{10}}
 \path(35,100)(165,100)
 \put(100,30){\circle*{10}}
 \put(100,170){\circle*{10}}
% \put(0,0){\scriptsize $p$}
% \put(190,0){\scriptsize $q$}
%
\thinlines
 \dashline[40]{10}(95,35)(35,95)
 \dashline[40]{10}(35,105)(95,165)
 \dashline[40]{10}(105,165)(165,105)
 \dashline[40]{10}(105,35)(165,95)
 \put(95,0){\scriptsize $\psi_S$}
 \put(95,190){\scriptsize $\psi_N$}
 \put(-5,95){\scriptsize $\rho_W$}
 \put(190,95){\scriptsize $\rho_E$}
% \put(95,80){\scriptsize $\theta$}
\end{picture}}
\end{picture}\caption{Types of faces containing white and black
 sites.}
\label{fig5}
\end{figure}
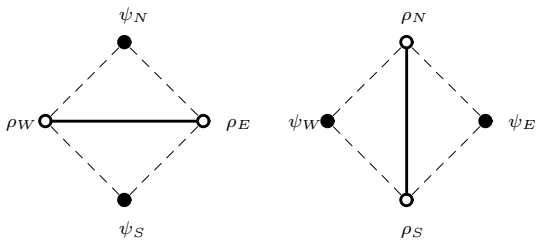

The cross ratio equations \eqref{cr-eq} are
closely related to the Hirota difference equation \cite{hirota3}.
Place a purely
imaginary variable $\psi_i$, \ $0\le {\rm Im}\,\psi_i<2\pi$, on every
site $i$ of the dual lattice.
These sites are shown
by black dots located in Figs.~\ref{fig4} and
\ref{fig5}. Connect these new variables to the existing
variables $\rho_i$ by the following rules.
Let $\psi_N$ and
$\psi_S$ be the variables located above and below of a horizontal edge,
as in Fig.~\ref{fig5}.
Then, we require that
\begin{equation}
\EXP^{\psi_N-\psi_S}=
\frac{\EXP^{\rho_W}+\EXP^{\rho_E+\ii\theta}}{\EXP^{\rho_E}
+\EXP^{\rho_W+\ii\theta}}\ .\label{hirota4}
\end{equation}
Similarly for a vertical edge,
\begin{equation}
\EXP^{\psi_E-\psi_W}=\frac{\EXP^{\rho_N}+\EXP^{\rho_S+\ii\theta^*}}{\EXP^{\rho_S}+\EXP^{\rho_N+\ii\theta^*}}
\Rightarrow \EXP^{\rho_N-\rho_S}=
\frac{\EXP^{\psi_W}+\EXP^{\psi_E+\ii\theta}}{\EXP^{\psi_E}+\EXP^{\psi_W+\ii\theta}}\;,\label{hirota5}
\end{equation}
where $\theta^*=\pi-\theta$.
The consistency of these definitions across the lattice is provided by
\eqref{cr-eq}.
Note that the second form of \eqref{hirota5} is identical to
\eqref{hirota4} upon interchanging all $\psi$- and $\rho$-variables.
Therefore it is natural to associate this universal
equation with every face of the ``double'' lattice consisting of all
white and black sites (its faces are shown in
Fig.~\ref{fig5}).
This is the famous Hirota equation introduced in \cite{hirota3}.

Thus, in the quasi-classical limit $\bs\to 0$ the
stationary points of the integral \eqref{Z-def} defines discrete
analogs of conformal transformations. At finite values of $\bs$ the
model describes quantum fluctuations of these transformations. Given
that the spins $\rho_i$ define the local dilatation factors the
Faddeev-Volkov model describes a {\em quantum discrete dilaton}.
The continuous quantum field theory with the conformal symmetry
\cite{BPZ84} has remarkable applications in physics and mathematics.
It would be interesting to understand which aspects of
the continuous conformal  field theory can be transferred to the
discrete case.

For the ``strongly coupled'' regime (ii), when the parameter $\bs$ is on
the unit circle $|\bs|=1$, the function
$W_{\frac{\pi}{2}}(s)$ still has its absolute maximum at
$s=0$ but starts
to develop additional side maxima at integer points when $\arg(\bs)$
approaches the value $\pi/2$.
\begin{figure}[htb]
\begin{center}
\includegraphics[scale=.7]{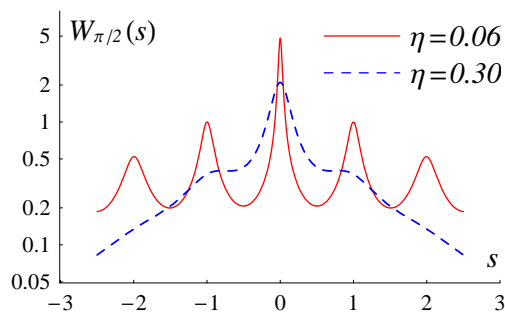}
\caption{The function $W_{\pi/2}(s)$ for $\eta=0.30$ and $0.06$.}
\label{fig6}
\end{center}
\end{figure}
The first pair of such side maxima at $s=\pm1$
appears when $\eta\simeq\cos(2\pi/5)\simeq0.309$, see Fig.~\ref{fig6}.
For $\eta=0$ the function
$W_{\frac{\pi}{2}}(s)$ turns  into a superposition of
$\delta$-function like peaks at $s\in{\mathbb Z}$.
In particular, the height of the central maximum diverges as
\begin{equation}
W_{\frac{\pi}{2}}(0)\simeq
(2\pi \crs)^{-1/2}\,\textstyle{\Gamma(\frac{1}{4})}/
{\Gamma(\frac{3}{4})}+O\big(\crs^{1/2}\big), \quad \eta\to 0\ .
\end{equation}
A more careful limiting procedure capturing a fine structure of these sharp
peaks requires a redefinition of the spin variables.
Using the asymptotics
\bea
\varphi(n+x\crs)&\simeq&\textstyle \EXP^{-\ii\pi/12} (4\pi\crs)^{\ii x}
{\Gamma(\frac{1-n+\ii x}{2})}/{\Gamma(\frac{1-n-\ii
x}{2})}\;,\nonumber\\[0.3cm]
\quad \Phi(2x\crs)&\simeq&\textstyle \EXP^{-\ii\pi/24}
(8\pi\crs)^{\ii x} {\Gamma(\frac{1+\ii
x}{2})}/{\Gamma(\frac{1-\ii x}{2})},
\eea
where $\eta\to0$, $n\in\mathbb{Z}$ and $|x|\ll \crs^{-1}$, it is easy
to see that
\begin{equation}\label{V}
W_\theta(n+x\crs)\;\simeq\; \crs^{-\theta/\pi}
V_\theta(n,x),\quad\eta\to 0
\end{equation}
where the function,
\begin{equation}
V_\theta(n,x)=\frac{1}{(2\pi)^{\frac{\theta}{\pi}}}
\frac{\Gamma(\frac{1+\theta/\pi}{2})\Gamma(\frac{1-n-\theta/\pi+\ii
x}{2})\Gamma(\frac{1-n-\theta/\pi-\ii
x}{2})}{\Gamma(\frac{1-\theta/\pi}{2})\Gamma(\frac{1-n+\theta/\pi+\ii
x}{2})\Gamma(\frac{1-n+\theta/\pi-\ii x}{2})}
\end{equation}
is real and positive for $n\in\mathbb{Z}$ and $x\in \mathbb{R}$.
This function defines Boltzmann weights for a new model where
each lattice site $i$ is assigned with a pair of fluctuating variables
$(n_i,x_i)$, where $n_i\in{\mathbb Z}$ take integer values
and $x_i\in\mathbb{R}$ take continuous values on the real line.
Its partition function is defined by \eqref{Z-def}
where $W_\theta(\sigma_i-\sigma_j)$ replaced by $V_\theta(n_i-n_j,x_i - x_j)$
(and similarly for $\overline{W}$) and every integral
over $\sigma_i$ is replaced by a sum over $n_i$ and an integral over
$x_i$, namely $\int d \sigma_i\to\sum_{n_i\in\mathbb{Z}}\int dx_i$.
\begin{widetext}
The star-triangle relation for $V_{\t\,}(n,x)$
simply follows from \eqref{str},
\begin{equation}
\begin{array}{l}
\ds \sum_{n_0\in\mathbb{Z}}\int_{\mathbb{R}} dx_0
V_{\theta_1}(n_1-n_0;x_1-x_0)V_{\theta_2}(n_2-n_0;x_2-x_0)
V_{\theta_3}(n_3-n_0;x_3-x_0)
\\
\phantom{V_{\theta_1}(n_1-n_0;x_1-x_0)V_{\theta_2}}
\ds=\;V_{\pi-\theta_1}(n_2-n_3;x_2-x_3)
V_{\pi-\theta_2}(n_1-n_3;x_1-x_3)
V_{\pi-\theta_3}(n_1-n_2;x_1-x_2)\;,
\end{array}
\end{equation}
where $\theta_1+\theta_2+\theta_3=2\pi$.
Similarly, the inversion relations \eqref{invrel}
imply
\begin{equation}
\lim_{\varepsilon\to 0^+} \sum_{n_2\in\mathbb{Z}}\int_{\mathbb R}
 dx_2 V_{\pi
+ \varepsilon}(n_1-n_2;x_1-x_2)V_{\pi -
\varepsilon}(n_2-n_3;x_2-x_3)
=\delta_{n_1,n_3}\delta(x_1-x_3)\;, \qquad V_\theta(n,x)V_{-\theta}(n,x)=1
\end{equation}
\end{widetext}
The Boltzmann weight $V_{\theta}(n;x)$ is an even function of $n$ and
$x$; it satisfies the following initial conditions
\bea
V_\pi(n;x)&=&\delta_{n,0}\delta(x)\;,\qquad V_0(n;x)=1\;.
\eea
Note that its asymptotics
when $n,x\to\pm\infty$,
\bea
V_\theta(n;x)&\simeq
&\Big(\frac{2}{\pi}\Big)^{\theta/\pi}\frac{\Gamma(\frac{1+\theta/\pi}{2})}
{\Gamma(\frac{1-\theta/\pi}{2})}\,
\big|\,n^2+x^2\,\big|^{-\theta/\pi},\qquad \;\;,\label{V-ass}
\eea
coincides with the {\em normalized} Boltzmann weight of the
$\mathscr{D}=2$ Za\-mo\-lod\-chi\-kov's ``fishing-net''
model \cite{Zamolodchikov:fishingnet}
(except that the variable $n$ in our case is discrete).
It follows from \eqref{fzero} that with
this normalization the specific free energy of
the ``fishing-net'' model vanishes in the thermodynamic
limit in complete agreement with
\cite{Zamolodchikov:fishingnet}.

\def\cprime{$'$} \def\cprime{$'$}

%\bibliography{total2}
%\bibliographystyle{vvb-bibstyle}
%\bibliographystyle{apsrev}

\end{document}